\documentstyle[aps,epsfig,preprint]{revtex}

\newcommand{\ba}{\begin{eqnarray}}
\newcommand{\ea}{\end{eqnarray}}
\newcommand{\bmath}{\begin{mathletters}}
\newcommand{\emath}{\end{mathletters}}
\newcommand{\ban}{\begin{eqnarray*}}
\newcommand{\ean}{\end{eqnarray*}}
\begin{document}

\tighten

\title{Extended M1 sum rule for excited symmetric and mixed-symmetry
states in nuclei}

\author{N. A. Smirnova$^{1,2}$\footnote{Nadya.Smirnova@fys.kuleuven.ac.be},
           N. Pietralla$^{3,4}$,
           A. Leviatan$^5$,
           J. N. Ginocchio$^6$,C. Fransen$^{4,7}$
            }

 \address{$^{1}$ Instituut voor Kern- en Stralingsfysica, University of
 Leuven, Celestijnenlaan 200 D, B-3001 Leuven (Heverlee), Belgium }
 \address{$^{2}$ ECT*, Str. delle Tabarelle, 286, 38050 Villazzano (Trento)
Italy }
 \address{$^{3}$ Wright Nuclear Structure Laboratory, Yale University,
 New Haven, CT, 06520, U.S.A.}
 \address{$^{4}$ Institut f\"ur Kernphysik, Universit\"at zu K\"oln,
        50937 K\"oln, Germany}
 \address{$^{5}$ Racah Institute of Physics, The Hebrew University,
 Jerusalem 91904, Israel}
 \address{$^{6}$ Theoretical Division, Los Alamos National Laboratory,
 Los Alamos, NM, 87545, U.S.A.}
 \address{$^{7}$ Physics Department, University of Kentucky,
 Lexington, KY, 40506, U.S.A.}
\date{\today}


\maketitle

\vspace{1cm}

\noindent Corresponding author: \\
Nadya A. Smirnova\\
University of Leuven, Instituut voor Kern- en Stralingsfysica,
Celestijnenlaan 200 D,
B-3001 Leuven (Heverlee), Belgium \\
Tel: (+32) 016 32 72 66 \\ Fax: (+32) 016 32 79 85 \\
E-mail: Nadya.Smirnova@fys.kuleuven.ac.be\\

\begin{abstract}
A generalized $M1$ sum rule for orbital magnetic dipole strength from
excited symmetric states to mixed-symmetry states is considered within the
proton-neutron interacting boson model of even-even nuclei. Analytic
expressions for the dominant terms in the $B(M1)$
transition rates from the $2^+_{1,2}$ states are derived in the U(5) and
SO(6) dynamic symmetry limits of the model, and the applicability of a sum
rule approach is examined at and in-between these limits. Lastly, the sum
rule is applied to the new data on mixed-symmetry states of $^{94}$Mo and
a quadrupole $d$-boson ratio $n_d(0^+_1)/n_d(2^+_2) \approx 0.6$ is obtained
in a largely parameter-independent way.\\
\end{abstract}


\newpage
\section{Introduction}

Heavy atomic nuclei exhibit both single-particle and collective excitations.
However, the coupling between these degrees of freedom can lead to strong
fragmentation of the collective modes. Under such circumstances sum rules
are useful, since they do not depend on the exact details of the
fragmentation and remain
applicable in cases where the collective modes
are not exact eigenstates of the Hamiltonian.
Sum rules generally express direct observables in terms of basic control
parameters ({\it e.g.} deformation)
which dominate the formation of the collective mode.
In cases where the relevant control parameter has natural boundaries,
one can obtain quantitative limits for observables in a largely
model-independent way.
Accordingly, sum rules are used both to judge what fraction of a
collective mode is present in a given ensemble of quantum states and as
a tool to exploit the link between direct observables and properties of
a given excitation mode.

One particular collective mode which has been studied extensively in
recent years is the orbital magnetic dipole scissors mode \cite{boh84},
which has by now been established experimentally as a general phenomenon
in nuclei \cite{ric95kne96}.
The systematics of $M1$ strength from the ground state to the
scissors state and its deformation dependence have been
extensively measured and corroborated with a variety of sum rules both in
even-mass \cite{gin91,heyde9194,zamick93,iudice93,enders99} and odd-mass
nuclei \cite{LeGi97}. Within the proton-neutron version of
the interacting boson model (IBM-2) \cite{ArOt77,OtAr78,IaAr87},
a sum rule \cite{gin91} has related this strength
to the number of quadrupole $d$-bosons in the ground state wave function.
For deformed nuclei the latter can be expressed in terms of the
quadrupole deformation determined from $B(E2)$ values, and the measured
$M1$ strength was shown to be in good agreement with this sum rule
\cite{vnc} as well as with its counterpart \cite{LeGi97} in odd-mass
nuclei \cite{huxel99}.

The measurement of the large transition rate between the mixed-symmetry ($\rm ms$) $1^+_{\rm 1,ms}$ and $2^+_{\rm 1, ms}$ states \cite{PiFr99} represents the first direct evidence
for the similar character of their wave functions. The discovery of the $3^+_{\rm 1,ms}$ mixed-symmetry state \cite{PiFr00} and of the $2^+_{\rm 2,ms}$ state \cite{FrPi01} on the
basis of electromagnetic transition matrix elements, added confidence in the general existence of mixed-symmetry states (even off-yrast) and allowed for the first time to judge
the energy splitting of the mixed-symmetry two-phonon quintuplet \cite{NPGoet}. This new data provides knowledge about $M1$ transition strengths from a set of mixed-symmetry
states \ba {\cal M} = \Bigl\{\,1^{+}_{\rm 1, ms}\;,\; 2^{+}_{\rm 1, ms}\;,\; 2^{+}_{\rm 2, ms}\;,\; 3^{+}_{\rm 1, ms}\,\Bigr\} \label{set} \ea to the symmetric $J=0^{+}_1$ ground
state {\em and} to the symmetric $J=2^{+}_{1},\;2^{+}_{2}$ excited states in $^{94}$Mo. This data can now be exploited in a new way, namely, the total $M1$ strengths between
mixed-symmetry states and different low-lying symmetric states of the same nucleus can be compared.

The empirical identification of the states in Eq.~(\ref{set})
relied on specific signatures as predicted by the IBM-2.
These IBM-2 predictions and assignments for mixed-symmetry states
were found to be in an impressive agreement with the data
\cite{PiFr99,PiFr00,FrPi01,FraPhD} and are further supported by
microscopic calculations \cite{sm,qpm}. This motivates us to use
the IBM-2 to extract structure information out of this extensive set of
measured $M1$ decay strengths in $^{94}$Mo.
In the present article we investigate how far a sum rule approach in the
IBM-2 may be used for that purpose, relying on the mixed symmetry
interpretation for the states in Eq.~(\ref{set}).
In section \ref{sec:sr} we present a sum rule for the $M1$
excitation strength from an arbitrary symmetric state which generalizes
an earlier expression for the total $M1$ ground state excitation strength
within the IBM-2.
Section \ref{sec:dsl} discusses the applicability of the sum rule
in the U(5) and SO(6) dynamic symmetry (DS) limits of the model as well as
in transitional cases preserving the SO(5) symmetry. In section
\ref{sec:Mo94} we apply the sum rule to the new data on $^{94}$Mo and
extract the relative quadrupole $d$-boson content of the $J=0^+_1$ and
$J=2^+_2$ states. The sum rule analysis is critically examined in
section V and the paper is summarized in section VI.

\section{Generalized $M1$ sum rule}
\label{sec:sr}
The standard one-body magnetic dipole ($M1$) operator in the IBM-2 has
the form~\cite{IaAr87}:
\begin{equation}
\label{TM1}
\hat{T}(M1)=\sqrt{\frac{3}{4\pi }}
\left(g_{\pi }\hat J_{\pi }+g_{\nu }\hat J_{\nu }\right)
= \sqrt{\frac{3}{4\pi }}\left[\,{(g_{\pi }+g_{\nu })\over 2}\hat J +
{(g_{\pi}-g_{\nu })\over 2}
\left(\hat J_{\pi }-\hat J_{\nu }\right)\, \right] ~,
 \end{equation}
where $\hat J_{\rho }$ are the individual boson angular momentum
operators for protons ($\rho=\pi$) or neutrons ($\rho=\nu$),
$g_{\rho}$ the respective boson g-factors and
$\hat J=\hat J_{\pi}+ \hat J_{\nu}$ the total angular momentum operator.
We are interested in a sum rule for the $M1$ strength from an excited state
in an even-even nucleus with angular momentum $J$ and maximal $F$-spin,
$F_{\rm max} = N/2 = (N_\pi +N_\nu)/2$.
The integers $N_{\rho}$ denote the total number
of proton or neutron bosons of monopole ($s$-) or quadrupole ($d$-) type.
These IBM-2 bosons represent correlated monopole and quadrupole
pairs of identical valence nucleons in the shell model \cite{ArOt77,OtAr78}.
The derivation of the sum rule follows the same steps as for the ground
state (which has $J=0^{+}$) given in Ref. \cite{gin91}, except that
the terms proportional to the total angular momentum are not dropped.
The derivation has been sketched already in Ref. \cite{LeGi97} in
which a sum rule for $M1$ ground state excitation strength of odd-mass
nuclei is derived.
The sum rule we are interested in here corresponds to the part due to the
core in Eq.~(6) of Ref.~\cite{LeGi97}. It is given by
 \begin{equation}
 \label{eq:sr}
 S_J = \sum_{f\neq i} B(M1;i,J \to f,J_f ) =
  6C\left[ \langle J | \hat{n}_d | J\rangle - \frac{J(J{+}1)}{6N}\right] \ ,
 \end{equation}
 where
 \begin{equation}
 \label{coeff}
 C ={3\over 4\pi}(g_{\pi} - g_{\nu})^2 {N_{\pi}N_{\nu}\over N(N-1)} ~.
 \end{equation}
Here $J$ ($J_f$) is the angular momentum of the initial (final) state
and the labels $i$ ($f$) indicate all quantum numbers that may be needed to
specify uniquely the states. In general $J\neq 0$ and, therefore, there can
be a magnetic dipole transition to the initial state proportional
to the magnetic moment. This elastic transition is not measured in the
reported experiments on $^{94}$Mo \cite{PiFr99,PiFr00,FrPi01,FraPhD},
hence, it is subtracted out in Eq.~(\ref{eq:sr}).
Since the IBM-2 states are assumed to have pure $F$-spin and the initial
state $J$ has $F=F_{\rm max}$, then only the
$\left(\hat J_{\pi }-\hat J_{\nu }\right)$ term in Eq.~(\ref{TM1})
(which is an F-spin vector) can contribute to $M1$ transitions,
and the sum in Eq.~(\ref{eq:sr}) involves all final
states subject to $F$-spin and
angular momentum selection rules: $F_{\rm max}\to F_{\rm max}-1$ and
$J_f = J-1, J, J+1$ ($M1$ transitions between states with $F=F_{\rm max}$
are forbidden due to the symmetry of their wave functions \cite{PVI86}).
The total $M1$ strength $S_J$ in Eq.~(\ref{eq:sr}) depends on the boson
numbers, $N_{\rho}$, the boson effective $g$-factors, $g_{\rho}$,
and involves the expectation value of the
$d$-boson number operator, $\hat{n}_d =\hat{n}_{d_\pi}+\hat{n}_{d_\nu}$
in the initial state $J$.
The dependence on $N_{\rho}$ reflects the local shell
structure and their values are fixed to be half the number of valence
particles or holes with respect to the nearest closed shell.
The boson $g$-factors defining the $M1$ operator of Eq.~(\ref{TM1})
are model-parameters which are needed
in order to extract from the sum rule information on
$n_d(J) = \langle J|\hat{n}_d |J\rangle$, the average number of
$d$-bosons in the IBM-2 wave function.
For the $J=0^{+}_1$ ground state, the sum rule in Eq.~(\ref{eq:sr}) reduces
to that of Ref.~\cite{gin91}. This special case was used earlier \cite{vnc}
to extract the $d$-boson content of the ground state from
the measured $M1$ excitation strengths. In that analysis the
parameters $g_{\rho}$ were assumed to have the values of bare orbital
$g$-factors, namely, $g_\pi = 1 \mu_N$ and $g_\nu = 0$.
The recent extensive measurements of $M1$
strengths in $^{94}$Mo \cite{PiFr99,PiFr00,FrPi01,FraPhD}
provides a way to avoid the assumption of effective boson $g$-factors by
considering ratios of $M1$ excitation strengths from different
symmetric states:
$R_{J_0}(J) = S_J/S_{J_0}$. Such ratios are independent of $g_{\rho}$
and are pure functions of the average numbers of $d$-bosons
in the states $J_0$ and $J$.
For example, if $J_0=2^{+}$ and $J=0^{+}_{1}$ it follows from
Eq.~(\ref{eq:sr}) that
\ba
\label{eq:Rsr}
R_{2^{+}}(0^{+}_{1}) &=& \frac{S_{0^{+}_{1}}}{S_{2^{+}}}
= \frac{n_d(0^{+}_{1})}{n_d(2^{+})-1/N}
\nonumber\\
&\approx&
\frac{n_d(0^{+}_{1})}{n_d(2^{+})} \quad
(\;n_d(2^{+}) \gg 1/N\;) ~ .
\ea
Thus, for $N$ sufficiently large, one can directly extract from the
measured $M1$ strengths, the relative $d$-boson contents of the
corresponding states in a largely parameter-independent way.
The relative $d$-boson content is
sensitive to the Hamiltonian parameters, {\it i.e.}, to the residual
interactions, and contains an important information on the structure of
wave functions.

Before a sum rule approach can be applied to the observed $M1$ strength
from low-lying symmetric states, it is crucial to
asses to what extent the mixed-symmetry states identified in $^{94}$Mo
can be expected to exhaust the sum rule.
For that purpose we need to examine the following partial
strengths $\Sigma_{J}$
\ba
 \Sigma_J = \sum_{f\in{\cal M}}
B(M1;i,J, F_{\rm max} \to f,J_f, F_{\rm max}{-}1)
\label{part}
\ea
to the set of mixed-symmetry states ${\cal M}$ of Eq.~(\ref{set}), and
compare to the full strengths $S_J$ in Eq.~(\ref{eq:sr}).
The analysis will be done first for the dynamic symmetries of the
IBM-2 and for transitional cases which are of relevance to $^{94}$Mo.
In the next section we consider specific types of Hamiltonians
in order to study the relative contributions to the sum rule
of different $M1$ branches from some low-lying states.

\section{M1 sum rules for F-spin invariant Hamiltonians}
\label{sec:dsl}

 In order to clarify the discussion we analyze in this section the
 contribution to the $M1$ sum rule from the two lowest lying
 $2^{+}$ states.
 Since we aim at the application of the sum rule to the
 $\gamma$-soft nucleus $^{94}$Mo, we pay particular attention
 to $F$-scalar Hamiltonians with SO(5) symmetry.
 We consider first the U(5) and SO(6) DS limits
 of the IBM-2 which contain the SO(5)
 subgroup, and derive analytic expressions for the relevant $M1$ excitation
 strengths (total strengths $S_J$ and partial strengths $\Sigma_J$) on top
 of the excited $2^+_1$ and $2^+_2$ states.
 Next we address the evolution of $B(M1;2^+_{1,2} \to J_f)$ values in a
 SO(5)-preserving transition path between the U(5) and SO(6)
 DS limits. The results from this section will serve as a guide-line
 for judging whether the currently available experimental data in
 $^{94}$Mo contains sufficient information to qualify for a sum rule
 analysis.

\subsection{U(5) and SO(6) DS limits}
\label{sub:U5}

 In the U(5) and SO(6) DS limits, the eigenstates of the Hamiltonian
have quantum numbers which are the labels of irreducible
representations (irreps) of the groups in the chains ~\cite{IaAr87},
\begin{equation}
\label{DSU5}
\begin{array}{ccccccccccc}
 \mbox{U}_{\pi }(6) & \otimes & \mbox{U}_{\nu }(6) & \supset &
\mbox{U}_{\pi \nu }(6) &
 \supset & \mbox{U}_{\pi \nu }(5) & \supset & \mbox{SO}_{\pi \nu }(5) &
\supset &
 \mbox{SO}_{\pi \nu }(3) \\
 \downarrow & & \downarrow & & \downarrow & & \downarrow  & &
 \downarrow & & \downarrow \\[0pt] [N_{\pi }] & \; &[N_{\nu }] & &
 [N_1,N_2] & & (n_1,n_2) & & (\tau_1,\tau_2) & \{\alpha_i\} & J \\
\end{array}
\end{equation}
 and
 \begin{equation}
 \label{DSO6}
 \begin{array}{ccccccccccc}
 \mbox{U}_{\pi }(6) & \otimes & \mbox{U}_{\nu }(6) & \supset &
\mbox{U}_{\pi \nu }(6) &
 \supset & \mbox{SO}_{\pi \nu }(6) & \supset & \mbox{SO}_{\pi \nu }(5) &
\supset &
 \mbox{SO}_{\pi \nu }(3) \\
 \downarrow & & \downarrow & & \downarrow & & \downarrow  & &
 \downarrow & & \downarrow \\[0pt] [N_{\pi }] & \; &[N_{\nu }] & &
[N_1,N_2] & &
 \langle \sigma_1,\sigma_2 \rangle & & (\tau_1,\tau_2) & \{\alpha_i\} & J \\
 \end{array}
 \end{equation}
respectively. Here $N_1 + N_2=N$, $F=(N_1-N_2)/2= N/2 -k$ $(k=0,1,\ldots)$,
while $\alpha_i $ ($i=1,2$) are missing labels,
necessary to completely classify the SO(5) $\supset $ SO(3) reduction.
Within these DS limits the average number of $d$-bosons in {\it any}
$F$-spin symmetric state is given by
 \begin{eqnarray}
{\rm U(5)}: \quad \quad n_{d}(F=F_{\rm max}, (n_d,0) ,(\tau, 0), J)
&=& n_d
\nonumber\\
{\rm SO(6)}: \quad n_{d}(F=F_{\rm max}, \langle \sigma =N,0\rangle ,
(\tau, 0), J) &=&
{N(N-1)\over 2(N+1)} + {\tau(\tau+3)\over 2(N+1)} ~. \label{u5o6}
\end{eqnarray}
For the lowest symmetric states with $J=0^{+}_1,\;2^{+}_{1},\;2^{+}_{2}$
and $\tau=0,1,2$, respectively, which are of particular interest to
the present discussion, we have
 \begin{eqnarray}
 \begin{array}{lcl}
 \displaystyle     & {\rm U(5)}\qquad & \qquad {\rm SO(6)} \\
 \displaystyle n_{d}(0^{+}_{1})\qquad & 0\qquad & {N(N-1)\over 2(N+1)} \\
 \displaystyle n_{d}(2^{+}_{1})\qquad & 1\qquad & {N(N-1)\over 2(N+1)}
+ {2\over N+1} \\
 \displaystyle n_{d}(2^{+}_{2})\qquad & 2\qquad & {N(N-1)\over 2(N+1)}
+ {5\over N+1} ~.
 \end{array} \label{nd}
 \end{eqnarray}
As can be seen, $n_{d}(2^{+})$ is of order unity in the U(5) limit and is
of order $N$ in the SO(6) limit. Therefore, the condition
$n_d(2^{+}) \gg 1/N$, mentioned in Eq.~(\ref{eq:Rsr}), is satisfied already
for $N \geq 4$ within $16\%$ for $J=2^{+}_1$ and
$11\%$ for $J=2^{+}_2$ near the SO(6) limit. Substituting the values of
$n_{d}(J)$ into Eq.~(\ref{eq:sr}) we obtain the following expressions for
the total $M1$ strength, $S_J$, from these states
 \begin{eqnarray}
 \begin{array}{lll}
 \displaystyle    &  \;\; {\rm U(5)} & \qquad {\rm SO(6)} \\
 \displaystyle S_{0^{+}_{1}}\qquad  & \;\;\; 0 \qquad & C\,
{3N(N-1)\over N+1} \\
 \displaystyle S_{2^{+}_{1}}        & C\,{6(N-1)\over N} &
C\,{3(N^2+2)(N-1)\over N(N+1)} \\
 \displaystyle S_{2^{+}_{2}}        &
C\,{6(2N-1)\over N} \qquad &
 C\,{3(N^3 - N^2 + 8N -2)\over N(N+1)} ~,
 \end{array}
 \label{srj}
 \end{eqnarray}
where $C$ is given in Eq.~(\ref{coeff}). The states which contribute to
these $M1$ strengths have the following classification in the U(5) or
SO(6) limits
 \ba
 \begin{array}{lcccc}
 & {\rm U(6)}\; F & {\rm U(5)} \;(n_1,n_2) \quad &
{\rm SO(6)}\; \langle\sigma_1,\sigma_2\rangle \quad &
 {\rm SO(5)} \; (\tau_1,\tau_2) \\[2mm]
 0^{+}_{1} \quad & F_{\rm max} \quad & (0,0) \quad &
\langle N,0\rangle\quad & (0,0) \\
 2^{+}_{1} \quad & F_{\rm max} \quad & (1,0) \quad &
\langle N,0\rangle\quad & (1,0) \\
 2^{+}_{2} \quad & F_{\rm max} \quad & (2,0) \quad &
\langle N,0\rangle\quad & (2,0) \\
 2^{+}_{\rm 1,ms} \quad & F_{\rm max} -1 \quad & (1,0) \quad & \langle
 N-1,1\rangle\quad & (1,0) \\ 1^{+}_{\rm 1,ms},\; 3^{+}_{\rm 1,ms} \quad &
 F_{\rm max} -1 \quad & (1,1) \quad & \langle N-1,1\rangle\quad & (1,1) \\
 2^{+}_{\rm 2,ms} \quad & F_{\rm max} -1 \quad & (2,0) \quad &
 \langle N-1,1\rangle\quad & (2,0) \\
 1^{+}_{\rm 2,ms},\; 2^{+}_{\rm 3,ms},\; 3^{+}_{\rm 2,ms} \quad &
F_{\rm max} -1 \quad &
 (2,1) \quad & \langle N-1,1\rangle \quad & (2,1) \\
 1^{+}_{\rm 3,ms},\; 2^{+}_{\rm 4,ms},\; 3^{+}_{\rm 3,ms},\;
 3^{+}_{\rm 4,ms} \quad & F_{\rm max} -1 \quad & (3,1) \quad &
 \langle N-1,1\rangle \quad & (3,1) \\
 \end{array}
 \label{u5o6st}
 \ea
The $M1$ operator of Eq.~(\ref{TM1}) transforms as a
$T^{[2,1^4]\{2,1^3\}(1,1)1 } $ tensor
under the U(5) chain, Eq.~(\ref{DSU5}), and as a
$T^{[2,1^4]\langle 1,1 \rangle (1,1)1}$
tensor under the SO(6) chain, Eq.~(\ref{DSO6}). Using standard techniques
for coupling irreps \cite{isa87} it is possible to show that
Eq.~(\ref{u5o6st}) lists {\it all} mixed-symmetry states which are relevant
for $M1$ transitions from the chosen initial states,
$J=0^{+}_1,\;2^{+}_1,\; 2^{+}_2$. The conservation of $d$-parity
\cite{WiGe97,d-parity} further restricts the allowed $M1$ transitions.
For SO(5) symmetry each state can be characterized by a definite value of
$d$-parity, $\pi_{d} =(-1)^{\tau_1+\tau_2}$, and the $M1$ operator has
$\pi_d=+1$. Altogether the only allowed $M1$ transitions
$J\;(\tau,0) \longrightarrow J_{f}\;(\tau_1,\tau_2)$ in the U(5) or
SO(6) DS limits are
 \bmath
 \ba
 \label{allowed}
 {\rm U(5)} & & \nonumber \\
 &&2^{+}_{1} \; (1,0)\longrightarrow 2^{+}_{\rm 1,ms} \; (1,0) \;\; , \;\;
\nonumber\\
 &&2^{+}_{2} \; (2,0) \longrightarrow 2^{+}_{\rm 2,ms} \; (2,0)\;\; ;  \quad
 1^{+}_{\rm 1,ms},\; 3^{+}_{\rm 1,ms} \; (1,1) \ ,\label{allowedu5} \\
 {\rm SO(6)} & & \nonumber \\
 &&0^{+}_{1} \; (0,0)\longrightarrow 1^{+}_{\rm 1,ms} \; (1,1) \;\; ,
\nonumber\\
 &&2^{+}_{1} \; (1,0)\longrightarrow 2^{+}_{\rm 1,ms} \; (1,0) \;\; ; \quad
 1^{+}_{\rm 2,ms},\; 2^{+}_{\rm 3,ms},\; 3^{+}_{\rm 2,ms} \; (2,1)\;\; ,
\nonumber\\
 &&2^{+}_{2} \; (2,0) \longrightarrow 2^{+}_{\rm 2,ms} \; (2,0)\;\; ; \quad
 1^{+}_{\rm 1,ms},\; 3^{+}_{\rm 1,ms} \; (1,1) \; ; \quad
1^{+}_{\rm 3,ms},\; 2^{+}_{\rm 4,ms},\; 3^{+}_{\rm 3,4,ms} \; (3,1) \ .
\label{allowedo6}
 \ea
 \emath
In the U(5) DS limit there are fewer allowed $M1$ transitions
due to an additional selection rule, namely, that $M1$ transitions can only
connect U(5) states $(n_1,n_2)$ such that $n_d=n_1+n_2$ is preserved.
Analytic expressions of $B(M1)$ values for $M1$ transitions in the U(5) and
SO(6) DS limits, which are relevant for the present discussion, are
collected in Table I.

In general, we see from Eq.~(13) that the total strengths, $S_J$, of
Eq.~(\ref{srj}) are sums of contributions involving different SO(5)
multiplets. Specifically, if we denote by $S_{J}^{(\tau_1,\tau_2)}$ the
summed $M1$ strength from the initial state $J$ to all final states with
$J_f=J,\,J\pm 1$ in a given SO(5) irrep $(\tau_1,\tau_2)$, we then find
 \bmath
 \ba
 S_{0^{+}_{1}} &=& S_{0^{+}_{1}}^{(1,1)} \\
 S_{2^{+}_{1}} &=& S_{2^{+}_{1}}^{(1,0)}+ S_{2^{+}_{1}}^{(2,1)} \\
 S_{2^{+}_{2}} &=& S_{2^{+}_{2}}^{(1,1)}+ S_{2^{+}_{2}}^{(2,0)}
+ S_{2^{+}_{2}}^{(3,1)}  ~.
 \label{srt}
 \ea
 \emath
From Table I we deduce the following expressions for the separate SO(5)
contributions to these strengths
 \ba
 \begin{array}{lcc}
  &  {\rm U(5)} & {\rm SO(6)} \\
 S_{0^{+}_{1}}^{(1,1)} \qquad &  0 \qquad & C\,{3N(N-1)\over N+1} \\
 S_{2^{+}_{1}}^{(1,0)} \qquad &  C\, {6(N-1)\over N} \qquad &
 C\, {3(N-1)(N+2)(N+4)\over 4N(N+1)}  \\
 S_{2^{+}_{1}}^{(2,1)} \qquad &
 0 \qquad & C\, {9(N-2)(N-1)\over 4(N+1)}  \\
 S_{2^{+}_{2}}^{(1,1)} \qquad &
 9C \qquad & C\, {9(N+4)(N+5)\over 14(N+1)}  \\
 S_{2^{+}_{2}}^{(2,0)} \qquad &
 C\, {3(N-2)\over N} \qquad & C\,{3(N+5)(N+2)(N-2)\over 10N(N+1)}  \\
 S_{2^{+}_{2}}^{(3,1)} \qquad &  0 \qquad &
C\, {72(N-2)(N-3)\over 35(N+1)} ~,
 \end{array}
 \label{partial}
 \ea
where $C$ is given in Eq.~(\ref{coeff}). As mentioned in section I, so far
only the mixed-symmetry states shown in Eq.~(\ref{set}), have been
identified in $^{94}$Mo.
Using their SO(5) classification given in Eq.~(\ref{u5o6st}),
we see that empirical information is available only for $M1$ strengths to
SO(5) multiplets with $(\tau_1,\tau_2)= (1,1),\; (1,0),\;(2,0)$.
No comparable firm information exists at present on the multiplets with
SO(5) quantum numbers $(2,1)$ and $(3,1)$. Considering the $M1$ strengths
that have been measured, the partial (yet observed) strengths, $\Sigma_{J}$,
of Eq.~(\ref{part}) can be transcribed as
\bmath
\ba \Sigma_{0^{+}_{1}} &=& S_{0^{+}_{1}}^{(1,1)} \label{srtp1} \\
\Sigma_{2^{+}_{1}} &=& S_{2^{+}_{1}}^{(1,0)} \label{srtp2} \\
\Sigma_{2^{+}_{2}} &=& S_{2^{+}_{2}}^{(1,1)}+ S_{2^{+}_{2}}^{(2,0)} ~.
\label{srtp3}
\ea
\emath
Comparing with the
corresponding expressions for total strengths in Eq.~(14), we see that the
$M1$ strength from the $J=0^{+}_1$ ground state arises entirely from the
transition $0^{+}_1\; (0,0) \to 1^{+}_{\rm 1,ms}\;(1,1)$, {\it i.e.},
$\Sigma_{0^{+}_{1}} = S_{0^{+}_{1}}$ in Eq.~(\ref{srtp1}). However, the
$M1$ transitions from the symmetric excited states, $2^{+}_1$ and $2^{+}_2$,
to the mixed-symmetry states of Eq.~(\ref{set}) exhaust only part of the
total strengths. Using Eq.~(\ref{partial}) these partial strengths in the
U(5) and SO(6) DS limits are found to be
 \ba
 \begin{array}{lcc}
  &  {\rm U(5)} & {\rm SO(6)} \\
 \Sigma_{2^{+}_{1}} \qquad &  C\, {6(N-1)\over N} \qquad &
 C\, {3(N-1)(N+2)(N+4)\over 4N(N+1)}  \\
 \Sigma_{2^{+}_{2}} \qquad &  C\, {6(2N-1)\over N} \qquad &
 C\, {3(N+5)(11N^2 + 30N - 14)\over 35N(N+1)} ~,
 \end{array}
 \label{u5o6p}
 \ea
where $C$ is given in Eq.~(\ref{coeff}).
In order to be able to apply the sum rule of Eq.~({\ref{eq:sr}) to the
existing data, we need to asses the goodness of the approximation
in replacing the total $M1$ strengths $S_{J}$ by the partial strengths
$\Sigma_{J}$. From Eqs.~(\ref{srj}) and (\ref{u5o6p}) we get the
following expressions for the ratios $Y(J)=\Sigma_{J}/S_{J}$
 \ba
 \begin{array}{lcc}
  &  {\rm U(5)} & {\rm SO(6)} \\
 Y(2^{+}_1)= \Sigma_{2^{+}_{1}}/S_{2^{+}_1} \qquad & 1 &
{(N+2)(N+4)\over 4(N^2+2)}  \\
 Y(2^{+}_2) = \Sigma_{2^{+}_{2}}/S_{2^{+}_2} \qquad & 1 &
 {(N+5)(11N^2 + 30N - 14)\over 35(N^3 - N^2 + 8N - 2)}~.
 \end{array}
 \label{ratio}
 \ea
We see that in the U(5) limit $Y(J)=1$ for $J=2^{+}_1,\;2^{+}_2$ and hence
the set ${\cal M}$ of mixed-symmetry states in Eq.~(\ref{set}) exhausts the
total $M1$ strength. This is not the case in the SO(6) limit in which
$Y(J) < 1$. In this case, the fraction $Y(J)$ of exhausted strength depends
on $N$ and is seen to be a monotonic decreasing function of
the boson number. For example, from Eq.~(\ref{ratio}) we have
$Y(2^{+}_1)=0.58,\; 0.41,\; 0.35$
and $Y(2^{+}_2)=0.85,\; 0.61,\; 0.51$ for $N=5,\; 10,\; 15$ respectively.
For large $N$ the asymptotic values are $Y(2^{+}_1)\rightarrow 0.25$ and
$Y(2^{+}_2)\rightarrow 0.31$.
We conclude that for nuclei near the SO(6) DS limit, the approximation
involved in the substitution $\Sigma_{J}\leftrightarrow S_{J}$ is better for
small $N$ and becomes less justified for large values of $N$.
To exhaust at least $75\%$ of the total strength
requires $N\leq 3$ for $J=2^{+}_1$ and $N\leq 6$ for $J=2^{+}_2$.
Furthermore, for a given $N$, $\Sigma_{2^{+}_{2}}$ is seen to provide a
better approximation to the total strength $S_{2^{+}_{2}}$, than
$\Sigma_{2^{+}_{1}}$ to $S_{2^{+}_{1}}$. This suggests that near
the SO(6) limit, a sum rule approach, based on the currently available data,
is likely to be reliable for moderate values of $N$ when applied to the
$J=0^{+}_1$ and $J=2^{+}_2$ states but not for the $J=2^{+}_1$ state.

\subsection{U(5) to SO(6) transition}
 \label{sub:U5toSO6}

The majority of transitional $\gamma$-soft nuclei lie in-between the U(5)
and SO(6) limits and retain good SO(5) symmetry. The main features of the
evolution in structure accompanying the transition between these two DS
limits can be studied by considering the following schematic $F$-scalar
Hamiltonian
 \begin{equation}
 \label{H}
 \hat H=a\left[(1-\zeta)\,\hat n_d
- \frac{\zeta}{4N}(\hat Q_{\pi }+\hat Q_{\nu })\cdot
 (\hat Q_{\pi }+\hat Q_{\nu }) + \lambda \hat M\right]\; .
 \end{equation}
Here $ \hat Q_{\rho }=[d^{\dagger}_{\rho } \times s_{\rho }
+ s^{\dagger}_{\rho } \times \tilde d_{\rho }]^{(2)} $ $(\rho=\pi,\nu)$
is the quadrupole operator relevant for this
transition region and $\hat M$ is the Majorana operator in Casimir form
\cite{IaAr87}. The Majorana term is diagonal and determines the energy shift
(proportional to $\lambda$) between eigenstates in accord with their
$F$-spin quantum numbers. Neither the parameter $\lambda$ nor the parameter
$a$ in Eq.~(\ref{H}) which sets the overall energy scale,
affect the structure of wave functions. The latter are completely determined
by the parameter $\zeta$ of $\hat H$. For $\zeta=0$ the Hamiltonian
possesses the U(5) DS, while for $\zeta=1$ it attains the SO(6) DS.
By varying $\zeta$ from $0$ to $1$ we can study in a simple
way the transition between the two limits. The calculations presented below
are done with $N_{\pi}=4$ and $N_{\nu}=1$ ($N=5$) which are the appropriate
boson numbers for $^{94}$Mo.

The top part of Fig.~1 shows the $d$-boson content, $n_d(J)$, of the
symmetric $J=0^+_1$, $2^+_1$, $2^+_2$ states with $F = F_{\rm max}$ as a
function of $\zeta$. The curves shown interpolate between the U(5) and SO(6)
values of Eq.~(\ref{nd}) for $N=5$. The lower part of Fig.~1 shows the
corresponding ratios of strengths, $R_{2^+_1}(J) = S_J/S_{2^{+}_1}$,
evaluated as in Eq.~(\ref{eq:sr}). For given boson numbers these ratios
depend only on the structural parameters of the Hamiltonian (in this case
only on $\zeta$) and not on parameters of the $M1$ operator in
Eq.~(\ref{TM1}). The sensitivity of such ratios to the transition path
between the U(5) and the SO(6) DS limits can be used to determine the
location of a given $\gamma$-soft nucleus along the transition leg between
these two DS limits.

The Hamiltonian of Eq.~(\ref{H}) is an F-scalar and although it does not have
a dynamic symmetry for arbitrary value of $\zeta$, it still always has an
SO(5) symmetry. Away from the U(5) and SO(6) DS limits, the eigenstates are
no longer pure with respect to U(5) nor SO(6). However, they do retain good
SO(5) quantum numbers and, consequently, the pattern of allowed $M1$
transitions shown in Eq.~(\ref{allowedo6}) persists also in the transition
region. In particular, the SO(5) and $d$-parity selection rules for $M1$
transitions are still in effect and the total strengths, $S_J$, maintain the
same SO(5) decomposition as in Eq.~(14). Fig.~2 displays the ratios of
partial to total strengths, $Y(J)=\Sigma_{J}/S_{J}$, as a function of
$\zeta$ for $J=0^{+}_1,\;2^{+}_1,\;2^{+}_2$. As shown, the partial strengths
$\Sigma_J$, to the set ${\cal M}$ of mixed symmetry states
of Eq.~(\ref{set}) exhaust the sum rules $S_{0^{+}_1}$ completely and
$S_{2^{+}_2}$ to a large extent (more than $85\%$) throughout the transition
region. Less than $15\%$ of the $M1$ strength from the $J=2^{+}_2$ state
goes into the SO(5) irrep $(3,1)$ which is not included in the partial
strength $\Sigma_{2^{+}_2}$ of Eq.~(\ref{srtp3}). On the other hand,
in most of the transition region, a considerable fraction of $M1$ strength
from the $J=2^{+}_1$ state is not concentrated in the above set of
mixed-symmetry states. About $40\%$ of the total strength $S_{2^{+}_1}$ goes
into the SO(5) irrep $(2,1)$ which is left out of the partial strength
$\Sigma_{2^{+}_1}$ in Eq.~(\ref{srtp2}). We conclude that
for $N=5$ throughout the transition region between the U(5) and SO(6) DS
limits, the partial strengths $\Sigma_J$ of Eq.~(\ref{part}) provide an
adequate approximation to the total strengths $S_J$ of Eq.~(\ref{eq:sr}) for
the $J=0^{+}_1$ and $J=2^{+}_2$ states but not for the $J=2^{+}_1$ state.
This identifies the initial states $J$ in $^{94}$Mo which qualify for a sum
rule analysis based on the measured $M1$ strengths to the mixed-symmetry of
Eq.~(\ref{set}).

\section{Application to $^{94}$Mo}
\label{sec:Mo94}

The primary goal of the present investigation is to extract structure
information, via a sum rule approach, out of
the recent extensive data on mixed symmetry states in $^{94}$Mo.
Table II displays a compilation of the available data on $M1$ transitions
from the $J=0^{+}_1,\;2^{+}_1,\;2^{+}_2$ states in $^{94}$Mo
\cite{PiFr99,PiFr00,FrPi01,FraPhD}. This data has been used
to identify the set ${\cal M}$ of mixed-symmetry states listed in
Eq.~(\ref{set}).
The experimental summed $M1$ strengths, $S(J)_{\rm Expt}$
given in Table II correspond to the calculated
partial strengths $\Sigma_J$ of Eq.~(\ref{part}) to these
mixed-symmetry states. In accord with the discussion of the previous
section (see in particular Fig.~2), for
a $\gamma$-soft nucleus such as $^{94}$Mo with $N=5$,
these partial strengths exhaust to a large extent
the $M1$ sum rule for $J=0^{+}_1$ and $J=2^{+}_2$.
For these states, it is therefore justified to
compare the measured ratio
\begin{equation}
\label{eq:Rexpt}
R_{2^+_2}(0^+_1)_{\rm Expt} =
\frac{S(0^+_1)_{\rm Expt}}{S(2^+_2)_{\rm Expt}}
 =  0.58^{+11}_{-14}
 \end{equation}
with the calculated ratio $S_{0^{+}}/S_{2^{+}_2}$ of total strengths
$S_J$ obtained from the sum rule in Eq.~(\ref{eq:sr}). Since in-between
the U(5) and SO(6) DS limits the value of $n_d(2^{+}_2)$ varies in the
range $2-2.5$ for $N=5$ [see Fig.~1 and Eq.~(\ref{nd})],
we can neglect $1/N=0.2$ with respect to $n_d(2^{+}_2)$ and, as in
Eq.~(\ref{eq:Rsr}), extract from the data a relative $d$-boson content
ratio
 \begin{equation}
 \left[\frac{n_d(0^+_1)}{n_d(2^+_2)}\right]_{^{94}{\rm Mo}} \approx
 0.58^{+11}_{-14}  ~.
\label{extract}
  \end{equation}
We find that the $J=0^{+}$ ground state of $^{94}$Mo contains more than
half as many $d$-bosons as the $J=2^+_2$ state. This number is considerably
higher than that for a spherical vibrator ($n_d(0^{+}_1)/n_d(2^{+}_2)=0$ in
the U(5) DS limit) and is in fact closer to the value expected for a
$\gamma$-unstable rotor
($n_d(0^{+}_1)/n_d(2^{+}_2)=2/3$ in the SO(6) DS limit with $N=5$).
These findings are consistent with previous observations that the $M1$ and
$E2$ strengths involving mixed-symmetry states in $^{94}$Mo compare favorably
with the SO(6) predictions \cite{PiFr99,PiFr00,FrPi01}. However, that
comparison relied on an assumption for the parameters of the $M1$ operator
(boson effective $g$-factors) and $E2$ operator (boson effective quadrupole
charges).
In the present approach such an assumption is avoided by using ratios of $M1$
strengths. The $d$-boson ratio of Eq.~(\ref{extract}) is extracted directly
from the data and its value is independent of any model parameters.

The $d$-boson content is sensitive to the transition path between the U(5)
and SO(6) limits, which are not easy to distinguish otherwise \cite{LeNo86}.
We can therefore use its empirical value to pin-down the location of
$^{94}$Mo along the transition leg in-between these limits. For that
purpose we show in Fig.~3 the calculated ratio $n_d(0^+_1)/n_d(2^+_2)$ as
a function of $\zeta$ (dashed line) and the value
$[n_d(0^+_1)/n_d(2^+_2)]_{^{94}{\rm Mo}} \approx 0.6$ of
Eq.~(\ref{extract}) extracted from the data (solid line). The comparison
between the calculated and empirical values strongly suggests a structural
parameter $\zeta > 0.7$ for the IBM-2 description of $^{94}$Mo and
unambiguously identifies this nucleus to be closer to the SO(6)
$\gamma$-unstable rotor rather than the U(5) spherical vibrator.

Besides the $M1$ properties, one may attempt to consider the known $E2$
rates in order to determine the appropriate parameter space of the IBM-2
Hamiltonian for $^{94}$Mo. An observable which can distinguish between the
U(5) and the SO(6) DS limits, is the shape invariant $K_4$
\cite{JoBr97,WePi00} which can be well approximated \cite{JoBr97} by the
experimentally accessible $B(E2)$ ratio
$K_4^{\rm appr.} = (7/10)\,B(E2;4^+_1 \to 2^+_1) / B(E2;2^+_1 \to 0^+_1)$.
For large $N$, $K_4=1.4$ in the U(5) limit and $K_4=1$ in the
$SO(6$) limit, with small deviations for finite $N$ \cite{WePi00}.
Unfortunately, for $^{94}$Mo the measured value \cite{BaBa72,Rama87} is
$K_4^{\rm appr.} = 1.16(17)$ and hence
the large error bars prohibit any definite conclusion about the symmetry
character of $^{94}$Mo from $E2$ data. More precise lifetime experiments on
low-lying symmetric states would be of interest for this issue.

\section{Critical Examination of the Analysis}
\label{sec:critical}

Some critical remarks on the implementation of the $M1$ sum rule are in
order. While Eq.~(\ref{eq:sr}) is an exact relation in the IBM-2, the
justification for applying it to the new data on $^{94}$Mo is less
straightforward and relies on the following assumptions.
(i)~All strong $M1$ transitions between low-lying states of $^{94}$Mo can
be modeled by the IBM-2.
(ii)~$F$-spin is a good symmetry for the states considered in $^{94}$Mo.
(iii)~The structure of $^{94}$Mo is consistent with a U(5)-to-SO(6)
transition path.

The first assumption is necessary to justify the comparison of the
experimental summed $M1$ strengths to the sums calculated in the IBM-2.
Sizeable, hypothetical contributions to the experimental
sums from states and degrees of freedom outside the IBM-2 space,
could potentially obscure the results.
However, the fact that $M1$ transitions in Table II
with strengths larger than $\approx 0.1 \mu_N^2$ are understandable in
the IBM-2, suggests that for $^{94}$Mo the excluded
degrees of freedom are not likely to have a significant impact on the empirical
summed strengths in the considered energy region. Futhermore, eventually
existing, additional strength can be
accounted for to a large extent by renormalizing the parameters of the
$M1$ operator in Eq.~(\ref{TM1}). In the present analysis we avoid any
assumption on these effective boson $g$-factors by considering ratios of
strengths.

The primary motivation for the use of the IBM-2 in the present sum rule analysis is the model's impressive success in predicting the experimental data in $^{94}$Mo
\cite{PiFr99,PiFr00,FrPi01,FraPhD}. The IBM-2 interpretation of these low-energy structures as mixed-symmetry states, implies that the $M1$ excitations involved are predominantly
of orbital character, and suggests that in $^{94}$Mo the spin contribution to the $M1$ strength at low energy is suppressed. This conjecture is supported by recent microscopic
studies of mixed-symmetry states in this nucleus \cite{sm,qpm}. Results of a realistic calculation within the quasiparticle-phonon model (QPM) \cite{qpm} indicate that
quantitatively, the $M1$ strengths resulting from using the standard values for the spin quenching factor ($g_s=0.6-0.7$) are larger than the experimental ones by at least a
factor of 2. The best overall agreement with experiments is reached for $g_s=0.3$. Even for $g_s=0.6$ the spin contribution is consistently smaller than the orbital one and is
about half  the orbital strength in the transition $1^{+}_{\rm 1, ms}\to 2^{+}_2$ for which the largest discrepancy between theory and experiment occurs. The low-lying spin
transitions were found to be very sensitive to small components of the wave functions, yet the appropriate quenching mechanism in the QPM has not been identified so far. A shell
model calculation for $^{94}$Mo \cite{sm} shows the isovector $M1$ ground state excitation strength to be concentrated in the $1^{+}_{\rm 1, ms}$ state and to be composed of
almost equal spin and orbital contributions. This considerable fraction of orbital $M1$ strength is to be regarded as a lower limit on the actual orbital contribution, given the
small model space used in the calculation ($^{88}$Sr core, employing large effective $E2$ charges), neglect of important orbitals, {\it e.g.}, $\pi(p_{3/2},g_{7/2})$, and
insisting on pure isovector $M1$ transitions. Clearly, large-scale shell model calculations are desirable to pin down the relative orbital and spin contributions to the $M1$
strength in $^{94}$Mo. Empirically, properties of the $J=1^{+}_{\rm 1, ms}$ state observed in $^{94}$Mo were found to be consistent with systematics of the scissors mode
extrapolated from the deformed rare earth nuclei \cite{PiFr99}. For the latter, the predominantly orbital character was empirically established by a comparison of
$(\gamma,\gamma^{\prime})$, $(e,e^{\prime})$ and $(p,p^{\prime})$ spectra \cite{wesselborg86}. It will be worthwhile in the future to verify experimentally to what extent such
dominance of orbital character for low-lying $M1$ strengths persists also in transitional nuclei such as $^{94}$Mo. This can be investigated by a comparison with inelastic
hadronic scattering and by exploiting the fact that while the orbital contribution is enhanced by deformation, the spin part has anti correlation with collectivity
\cite{auerbach93}.

The second assumption of good $F$-spin symmetry is the basis for the
derivation of the sum rule in Eq.~(\ref{eq:sr}). Various procedures have
been proposed to estimate the $F$-spin purity of low-lying states in nuclei
\cite{lipas90}. These involve examining $M1$ transitions (which should
vanish between pure $F=F_{\rm max}$ states \cite{PVI86}),
magnetic moments \cite{gino92,wolf93}, the difference in proton-neutron
deformations \cite{lgk90} and properties of $F$-spin multiplets
\cite{harter85,lev00}. In the majority of
analyses the $F$-spin admixtures in low-lying states are found to be a few
percent ($< 10\%$) typically $2-4\%$ \cite{lipas90}. Although the empirical
$M1$ strengths shown in Table II are fragmented, the pattern of dominant
transitions to the mixed symmetry states in $^{94}$Mo as well as their
energy systematics agree favorably with the assignment of $F$-spin quantum
numbers. The smallness of the observed $M1$ rate,
$B(M1;2^{+}_2 \to 2^{+}_1) = 0.06(2)\ \mu_N$, which is $F$-spin forbidden,
is a benchmark for the anticipated $F$-spin mixing in low-lying states of
$^{94}$Mo.

The last assumption is adequate for $\gamma$-soft nuclei and ensures that
the SO(5) symmetry is preserved. This additional symmetry played a
significant role in the current analysis by imposing further constraints on
the allowed $M1$ transitions, which in turn enabled the observed four
mixed-symmetry states of Eq.~(\ref{set}) to exhaust an appreciable fraction
of the sum rules, $S_J$, for $J=0^{+}_1$, $2^{+}_2$.

\section{Summary}

The recent extensive data in $^{94}$Mo \cite{PiFr99,PiFr00,FrPi01,FraPhD}
on the four mixed-symmetry states listed in Eq.~(\ref{set}), has paved the
way for comparing $M1$ strengths between mixed-symmetry states and
different low-lying symmetric states in the same nucleus.
Sum rules are the proper tool to study the systematics of total strengths
in the presence of fragmentation.
The success of the IBM-2 in reproducing the data
has motivated us to consider a generalized
$M1$ sum rule from any symmetric state in the IBM-2 framework.
The sum rule is a generalization to excited states of an earlier
sum rule for the ground state and it relates the total $M1$ excitation
strength to the average number of $d$-bosons,
$n_d(J)$, in the IBM-2 wave function of the initial state $J$.
The latter is an important quantity characterizing the state and is
linked with its deformation.
By applying the sum rule to different initial
states and taking ratios of the total strengths, one can
avoid any assumption on the effective-boson $g$-factors and thus
eliminate to a large extent a model-dependence from the extracted ratios of
$d$-boson contents.

Before the sum rule can be applied, one needs, however, to be sure that the
experimental summed $M1$ strengths to the mixed-symmetry states of
Eq.~(\ref{set}) exhaust a significant fraction of the total $M1$ strengths.
This was verified to be the case,
analytically, for the U(5) and SO(6) DS limits and, numerically, throughout
the transition region in-between these limits. The analysis employed
$F$-spin scalar and SO(5) invariant Hamiltonians relevant for $\gamma$-soft
nuclei. The presence of an additional SO(5) symmetry restricts the allowed
$M1$ transitions and for $N=5$ enables the mixed-symmetry
states of Eq.~(\ref{set}) to exhaust more than $85\%$ of the sum rule for
the $J=0^{+}_1$ and $2^{+}_2$ states. We have applied the sum rule to
$^{94}$Mo and deduced from the data a relative $d$-boson content ratio
$n_d(0^+_1)/n_d(2^+_2)  \approx 0.6$. The extracted value is independent of
any model-parameters and suggests the structure of $^{94}$Mo being close to
the SO(6) DS limit of the IBM-2. The results obtained show that existing and
future high-quality data on excited mixed-symmetry states in nuclei can
qualify for a sum rule analysis from which one can extract valuable
model-independent structure information.
The present analysis relies on the IBM-2 interpretation of mixed symmetry
states as predominantly orbital $M1$ excitations. This interpretation
is consistent with presently available data in $^{94}$Mo
\cite{PiFr99,PiFr00,FrPi01,FraPhD} and with microscopic calculations
\cite{sm,qpm}. Further theoretical and experimental work on orbital and
spin excitations are highly desirable to verify the character of $M1$
excitations in transitional nuclei such as $^{94}$Mo.

 \acknowledgements{ N.A.S. and N.P. are indebted to P.~Van~Isacker for helpful discussions and
 gratefully acknowledge discussions with P.~von Brentano, R.~F.~Casten,
 A.~E.~L.~Dieperink, F.~Iachello, J.~Jolie, E.~Lipparini, N. Lo\,Iudice, T.~Otsuka,
 S.W. Yates and N.~V.~Zamfir. N.A.S. is a post-doctoral researcher of the FWO-Vlaanderen,
 Belgium
 and is grateful for the hospitality received at the European Centre of Theoretical Physics
 and Related Areas (Trento) during a stay in which a part of this work was done.
 The authors thank
 the Institute for Nuclear Theory at the University of Washington for its hospitality.
 This work is supported in part by a Grant-in-Aid for Scientific Research (B)(2)(10044059)
 from the Ministry of Education, Science and Culture,
 by the U.S. DOE under grant No. DE-FG02-91ER-40609 and contract W-7405-ENG-36,
 by the Deutsche Forschungsgemeinschaft under
 grant Nos. Pi 393/1-1/1-2 and by a grant from the Israel Science Foundation.}

\begin{table}
\label{tab:emSO6}
\caption{Some relevant analytic expressions for $B(M1)$ values in the U(5)
and SO(6) DS limits for $M1$ transitions from symmetric states
\protect$(F=F_{\rm max})$ to mixed-symmetry states
\protect$(F = F_{\rm max} - 1)$ and SO(5) quantum numbers
$(\tau_1,\tau_2)$ as indicated. The factor $C$ is given in
Eq.~(\protect\ref{coeff}).}
\begin{tabular}{lll}
Transition & U(5)$^a$ & SO(6) \\
\hline $0^{+}_1\; (0,0) \to 1^{+}_{\rm 1,ms}\; (1,1)$ & $0$ &
$C\, {3N(N-1)\over N+1}\;$ $^b$ \\
$2^{+}_1\; (1,0) \to 2^{+}_{\rm 1,ms}\; (1,0)$ &
$C\,{6(N-1)\over N} $ & $C\,{3(N+4)(N+2)(N-1)\over 4N (N+1)}\;$ $^b$ \\
$2^{+}_1 \; (1,0) \to 1^{+}_{\rm 2,ms}\; (2,1)$ &
$0$ & $C\,{3(N-1)(N-2)\over 10 (N+1)}$ \\
$2^{+}_1 \; (1,0) \to 2^{+}_{\rm 3,ms}\; (2,1)$&
$0$ & $C\,{3(N-1)(N-2)\over 4 (N+1)}$ \\
$2^{+}_1 \; (1,0) \to 3^{+}_{\rm 2,ms}\; (2,1)$& $0$ &
$C\,{6(N-1)(N-2)\over 5 (N+1)}$ \\
$2^{+}_2 \; (2,0) \to 1^{+}_{\rm 1,ms}\; (1,1)$& $C\,{21\over 5}$ &
$C\,{3 (N+5)(N+4)\over 10 (N+1)}\;$
$^b$ \\ $2^{+}_2 \; (2,0) \to 3^{+}_{\rm 1,ms}\; (1,1)$&
$C\,{24\over 5}$ & $C\,{12(N+5)(N+4)\over 35 (N+1)}$\\
$2^{+}_2 \; (2,0) \to 2^{+}_{\rm 2,ms}\; (2,0)$&
$C\,{3(N-2)\over N}$ & $C\,{3 (N+5)(N+2)(N-2)\over 10 N(N+1)}$
\end{tabular}
$^a$ from Ref \cite{Gian93}\\
$^b$ from Ref \cite{PVI86}
\end{table}

{\tighten

\begin{table}

\caption{Measured $M1$ transition strengths in \protect$^{94}$Mo in
units of $\mu_N^2$ \protect\cite{PiFr99,PiFr00,FrPi01,FraPhD}.
The notation ``n.o.'' denotes cases where the corresponding transitions
were too weak to be observed, although other decay branches of the issuing
level were detected. The states $1^{+}_1$, $3^{+}_2$, $2^{+}_3$, $2^{+}_6$
are the main fragments of the $1^{+}_{\rm 1,ms}$, $3^{+}_{\rm 1,ms}$,
$2^{+}_{\rm 1,ms}$ and $2^{+}_{\rm 2,ms}$ mixed-symmetry states
respectively. $S(J)_{\rm Expt}$ is the experimental summed $M1$
strength from the initial state $J$.}
\smallskip
\begin{tabular}{ld}
Observable & Expt. \\
$B(M1;0^+_1 \rightarrow 1^+_1)$  &    0.47(3)\tablenote{from Ref
\cite{PiFr99}} \\
 $B(M1;0^+_1 \rightarrow 1^+_2)$  &    0.14(5)$^{\rm a}$  \\
 \hline
$S(0^+_1)_{\rm Expt}$                       &    0.61(7)$^{\rm a}$  \\[2mm]

$B(M1;2^+_1 \rightarrow 1^+_1)$  &    0.004$^{+4}_{-1}$$^{\rm a}$   \\
$B(M1;2^+_1 \rightarrow 1^+_2)$  &    0.007(4)$^{\rm a,d}$  \\

$B(M1;2^+_1 \rightarrow 2^+_2)$  &    0.06(2)$^{\rm a}$   \\
$B(M1;2^+_1 \rightarrow 2^+_3)$  &    0.48(6)$^{\rm a}$    \\
$B(M1;2^+_1 \rightarrow 2^+_4)$  &    0.07(2)$^{\rm a}$    \\
$B(M1;2^+_1 \rightarrow 2^+_5)$  &    0.03(1)$^{\rm a}$    \\
$B(M1;2^+_1 \rightarrow 2^+_6)$  & $<$0.0077\tablenote{from Ref
\cite{FrPi01}}   \\
$B(M1;2^+_1 \rightarrow 3^+_2)$  & 0.014$^{+17}_{-8}$\tablenote{from Ref
\cite{PiFr00}}    \\
\hline
$S(2^+_1)_{\rm Expt}$                       &    0.67(7)  \\[2mm]

$B(M1;2^+_2 \rightarrow 1^+_1)$  &    0.26(3)$^{\rm a}$   \\
$B(M1;2^+_2 \rightarrow 1^+_2)$  &    n.o.\tablenote{from Ref
\cite{FraPhD}}      \\

$B(M1;2^+_2 \rightarrow 2^+_3)$  &    n.o.$^{\rm d}$     \\
$B(M1;2^+_2 \rightarrow 2^+_4)$  & $<$0.02$^{\rm b}$     \\
$B(M1;2^+_2 \rightarrow 2^+_5)$  &    0.095(6)$^{\rm b}$  \\
$B(M1;2^+_2 \rightarrow 2^+_6)$  &    0.35(11)$^{\rm b}$  \\
$B(M1;2^+_2 \rightarrow 2^+_7)$  &    0.009$^{+7}_{-3}$$^{\rm b}$    \\
$B(M1;2^+_2 \rightarrow 2^+_8)$  &    n.o.$^{\rm d}$   \\

 $B(M1;2^+_2 \rightarrow 3^+_2)$  &    0.34$^{+20}_{-10}$$^{\rm c}$    \\
 \hline
$S(2^+_2)_{\rm Expt}$                       &    1.05$^{+23}_{-15}$  \\

\end{tabular}
\label{tab:data}
\end{table}}

\begin{figure}
\label{fig:sum}
\epsfig{file=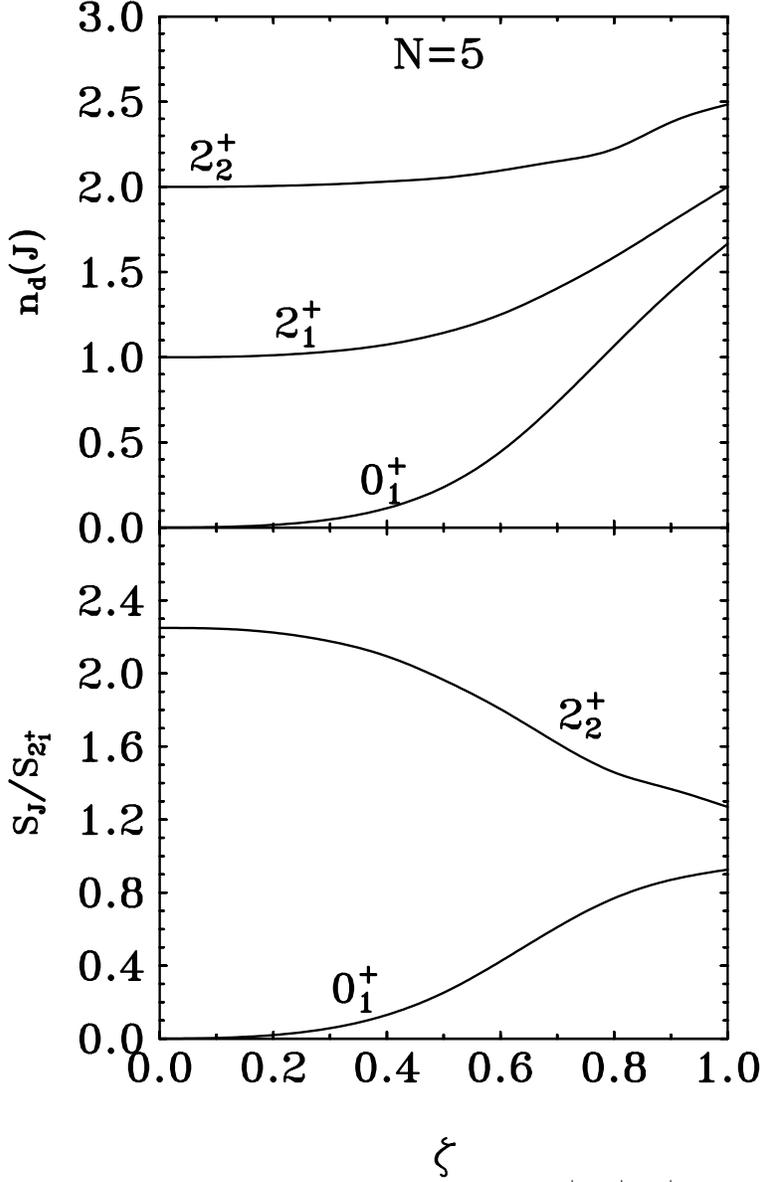,width=10cm}
\caption{$d$-boson content, $n_d(J)$, of the $J=0^+_1$,
$2^+_1$, $2^+_2$ states (top) and the corresponding ratios of total
\protect$M1$ excitation strengths $S_{0^+_1}/S_{2^+_1}$ and
$S_{2^+_2}/S_{2^+_1}$ (bottom) as a function of
\protect$\zeta$. Calculations are done with the Hamiltonian of
Eq.~$(\protect\ref{H})$ with boson numbers $N_{\pi }=4$ and $N_{\nu }=1$
$(N=5)$.}
\end{figure}

\begin{figure}
\label{fig:YJoverSJ}
\epsfig{file=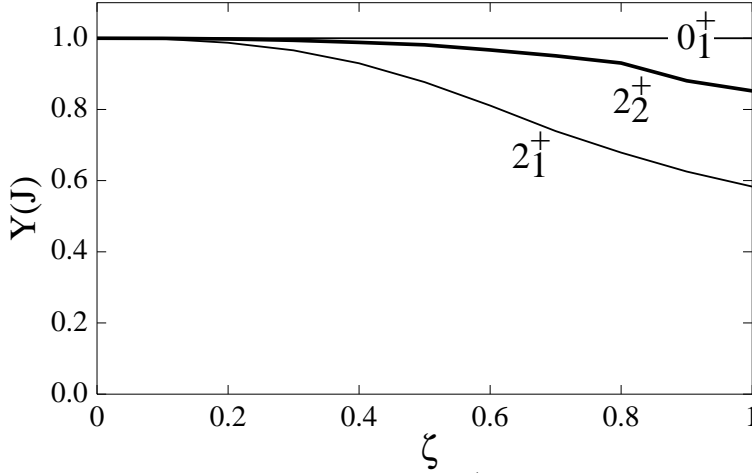,width=10cm}
\caption{Calculated ratios $Y=\Sigma_J/S_J$ of partial
to total $M1$ strengths as a function of $\zeta$ for
$J=0^{+}_1,\;2^{+}_1,\; 2^{+}_2$ and
$N=5$. The partial strengths $\Sigma_J$ to the mixed-symmetry states of
Eq.~(\protect\ref{set}) exhaust the sum rules of Eq.~(\protect\ref{eq:sr}),
\protect$S_{0^+_1}$ completely,
and \protect$S_{2^+_2}$ to more than $85\%$ in the whole U(5)-to-SO(6)
transition path.}
\end{figure}

\begin{figure}
\label{fig:nd01dnd22vszeta}
\epsfig{file=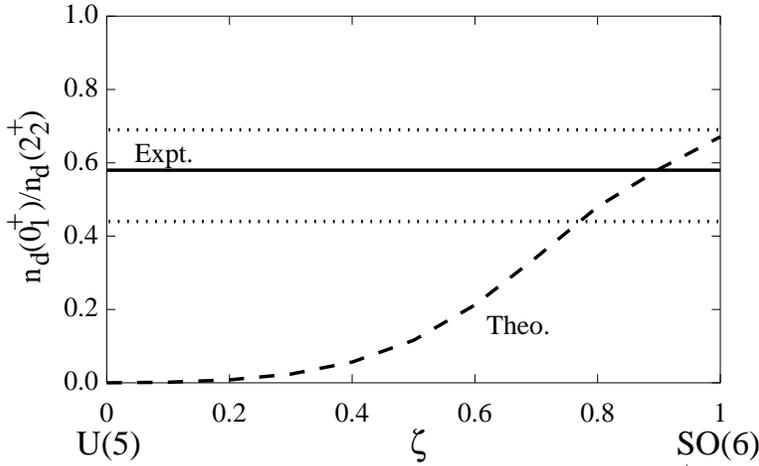,width=10cm}
\caption{The calculated $d$-boson ratio of the $J=0^+_1$ and $J=2^+_2$
states (dashed curve) as a function of $\zeta$ compared with the empirical
value (solid line with experimental uncertainties indicated by the dotted
lines) of Eq.~(\protect\ref{extract}) extracted from the measured $M1$
strengths in $^{94}$Mo. }
\end{figure}

\end{document}